# Reducing Thermodynamics to Boltzmannian Statistical Mechanics: The Case of Macro Values

Alexander Ehmann

alexanderehmann@alexanderehmann.com

Abstract:

Thermodynamic macro variables, such as the temperature or volume macro variable, can take on a continuum of allowable values, called thermodynamic macro values. Although referring to the same macro phenomena, the macro variables of Boltzmannian Statistical Mechanics (BSM) differ from thermodynamic macro variables in an important respect: within the framework of Boltzmannian Statistical Mechanics the evolution of macro values of systems with finite available phase space is invariably modelled as discontinuous, due to the method of partitioning phase space into macro regions with sharp, fixed boundaries. Conceptually, this is at odds with the continuous evolution of macro values as described by thermodynamics, as well as with the continuous evolution of the micro state assumed in BSM. This discrepancy I call the discontinuity problem (DP). I show how it arises from BSM's framework and demonstrate its consequences, in particular for the foundational project of reducing thermodynamics to BSM: thermodynamic macro values are shown to not supervene on the corresponding BSM macro values. With supervenience being a *conditio sine qua non* for the kind of reduction envisaged by the foundational project, the latter is in jeopardy.

Introduction

A core method of the framework of Boltzmannian Statistical Mechanics (BSM) is the partitioning of phase space into non-overlapping macro regions, i.e., disjoint sets of micro states. Macro regions are of finite, non-zero volume, and stand in a one-to-one correspondence with different macro states.[1] Given a finite volume of phase space to be partitioned, there is a finite number of macro regions in a partition. Macro states are specified via particular macro values, e.g., for temperature, volume and pressure of a gas. In general, macro states are defined and specified via particular macro values taken on by macro variables: following Werndl & Frigg (2015, 1225), a system "can be characterized by a set $\{v_1, \dots, v_k\}$ of *macrovariables* ($k \in \mathbb{N}$). The $v_i$ assume values in [macro value space] $\mathbb{V}_i$, and capital letters $V_i$ denote the values of $v_i$. A particular set of values $\{V_1, \dots, V_k\}$ defines a *macrostate $M_{V_1,\dots,V_k}$*."

Boundaries between macro regions are construed as being sharp: under a given partition, each micro state is a member of one and only one macro region. Since macro regions stand in a one-to-one correspondence with macro states – i.e., there is a bijection between the set of macro regions and the set of macro states – each micro state instantiates one and only one macro state, and micro states from different macro regions instantiate different macro states. Also, once fixed, a given partition doesn't change (Hemmo & Shenker 2012, 53): "The partition of the state space into macrostate regions is fixed in time"; and (ibid., 56): "the partition of the state space into macrostates […] is time-independent". This partition of phase space into a finite number of macro regions with sharp, fixed boundaries results in models that are unable to describe continuous macroscopic change: whenever macroscopic change is modelled within the framework of BSM, it is modelled as discontinuous. That is, according to BSM's models, every time the micro state of a system crosses the boundary between two macro regions, a discontinuity in (at least one of) the system's macro values results.

---

[1] See, e.g., Callender (1999), Frigg & Werndl (2012), Werndl & Frigg (2015), Goldstein et al. (2017). The following discussion is based mainly on the framework of BSM as presented in the writings of Charlotte Werndl and Roman Frigg, but also others. I will often refer to it as the "usual framework", for it is fairly common, without intending to imply that there aren't any alternatives. Thus, the issues identified pertain at least to this "usual" account, but not necessarily to all possible accounts. In particular, I deem it possible to avoid the discontinuity problem described below by adopting a different way of carving up phase space. Such an alternative I will present in a companion paper.

Conceptually, this is at odds with the thermodynamic description of the continuous time-evolution of macro values, and likewise with the continuous evolution of the micro state in phase space. From this discontinuity problem (DP), as I call it, ensues a problem for the foundational project[2] of reducing thermodynamics to BSM.[3] The foundational project is, as Callender (1999, 348) puts it, "the project of demonstrating how real mechanical systems can behave thermodynamically."

Take, for example, a gas consisting of a number $N$ of particles and assume that classical mechanics provides a correct description of the fundamental goings-on, i.e., of the dynamics of the $N$ particles. The micro state of such a $N$-particle system is represented by a point $x$ in $6N$-dimensional phase space $\Gamma$, specifying all positions and momenta of the $N$ particles. "As the system evolves through time, this representative point will trace out a trajectory through $\Gamma$. We thus have two descriptions of our gas: one mechanical and the other thermodynamical. We would like to know how they relate to one another. This kind of problem is quite familiar to philosophers. Philosophers of biology, psychology, and all of the special sciences are busy trying to demonstrate how the properties and concepts used by their science are (not) reducible in some sense to the properties and concepts of 'lower level' sciences. The problem is one of intertheoretic reduction." (Callender 1999, 350f.)

The thrust of this paper is to show that this inter-theoretic reduction of thermodynamics to BSM isn't as straightforward as it might seem at first – even in the seemingly simple case of reducing the thermodynamic concept of temperature to its analogue in BSM – and to show why this is so. This problem will be framed in terms of models and their (in)adequate tracking of their target systems. The question to be answered, then, is the following: Is the evolution of macro values of thermodynamic target systems adequately tracked by BSM's models? It will be answered in the negative: the BSM framework, in principle, doesn't allow for models that adequately track the evolution of thermodynamic macro values. This, while

---

[2] I adopt the terminology from Callender (1999, 348).
[3] Throughout this paper, I adopt the "Philosopher's sense of reduction", see Batterman (2020).

being unproblematic for the practitioner, constitutes a conceptual hurdle for the foundational project.[4]

Many of the literature on the reduction of thermodynamics to statistical mechanics engages with the tension between time-reversal invariant micro dynamics and the second law of thermodynamics. This is not the subject of the present investigation. The problem at hand, in a sense, is more basic: while thermodynamic descriptions allow for continuous change of temperature macro values, descriptions provided by BSM's models don't. It is in this sense that the latter do not adequately track the former. The two descriptions come apart: macroscopic temperature, as described by BSM, behaves differently in its evolution over time than macroscopic temperature as described by thermodynamics. This indeed is puzzling, and poses a threat to the foundational project: with macroscopic temperature in thermodynamics ($T_{TD}$) differing in this way from macroscopic temperature in BSM ($T_{BSM}$), the possibility of reducing one to the other is put into question. At the very least, $T_{TD}$ and $T_{BSM}$ aren't readily identified with one another, as might seem possible at first sight. What's more, as will become evident later, $T_{TD}$ doesn't supervene on $T_{BSM}$, a condition generally regarded as a necessary requirement for reduction. As van Riel & Van Gulick (2019, 4.5.3) write: "it [supervenience] surely is a necessary condition for reduction". Likewise, McLaughlin & Bennett (2018, 3.3): "Everyone agrees that reduction requires supervenience". But with supervenience being a *conditio sine qua non* for reduction, and $T_{TD}$ not supervening on $T_{BSM}$, the former cannot be reduced to the latter.

The aim of this paper is *not* to argue against the foundational project of reducing thermodynamics to BSM, or to attack the latter. Rather, the goal is to contribute to the foundational project by providing counterexamples that show where BSM and its models, as it stands, have their shortcomings, and how these threaten the foundational project. Following Wimsatt (2006, 473), "counterexamples become revealing sources of information about limitations of a model, or suggestions for probing its depths; in either case, a tool to refine the model, not an argument for trashing the system, or something to be swept under the rug."

---

[4] Like Callender's (see ibid., 349), my quarrels are philosophical in nature, concerning the foundational project. The practicing scientist needn't be concerned.

Models describing change of macro values as always discontinuous are at odds with thermodynamic descriptions of these macro values: a system's macroscopic evolution as described by thermodynamics involves continuous change of its macro properties – of the temperature value, for example. Likewise, such models are at odds with the continuous evolution of the micro state $x$ through $\Gamma$. A continuous evolution of the micro state should not lead to discontinuous change of the macro state, i.e., should not lead to a discontinuity in the evolution of macro values. Yet, as the framework of BSM is set up, continuous change of the micro state sometimes (i.e. upon traversing boundaries between macro regions) *does* lead to discontinuous change of the macro state. In fact, the models resulting from the framework of BSM predict that, *whenever* there is macroscopic change, it is sudden and of a jumpy nature: during its wanderings *within* a macro region, the micro state always and invariably instantiates the same, unchanging macro state, according to the framework's models. As soon as the micro state moves into a *different* macro region, however, it suddenly instantiates a different macro state: since boundaries between macro regions are construed as sharp, macro values jump upon traversal of the micro state from one macro region to another. This results from the definition of macro states via "particular set[s] of [macro] values", and the bijective association of the so defined macro states with macro regions of finite volume (Werndl & Frigg 2015, 1225): "a system's microstate uniquely determines its macrostate. Every macrostate $M$ is associated with a macroregion $\Gamma_M$ consisting of all $x \in \Gamma_E$ [with $\Gamma_E$ denoting the energy hypersurface, i.e. the complete available phase space for a constant energy] for which the system is in $M$." The consequence of this construal is that at least one macro value always jumps upon the micro state crossing boundaries between macro regions. Assuming that the continuous thermodynamic evolution of macro states is to be recovered when reducing thermodynamics to BSM, jumping macro values, especially upon tiny changes of the micro state, are ill-suited. Yet, the partition of phase space into disjoint macro regions of finite, non-zero volume with fixed, sharp boundaries, as prescribed by the usual framework, results in models describing jumping macro states.

This discontinuity problem, merely sketched so far, will be discussed in detail in the following. At this point, some emphasis should be put on the fact that DP only pertains to models resulting from the usual framework of BSM, and explicitly not to the target systems described by those models. The claim is neither that the physical systems themselves suffer from DP, i.e., that they always undergo discontinuous macroscopic change, if they undergo macroscopic change at all. Nor, that the thermodynamic description of these physical systems suffers from DP. Rather, only the BSM models that describe them entail these unavoidable discontinuities. This, in itself, would even be unproblematic. Per se, there is nothing wrong about models allowing only for discontinuous change. DP arises only if one wants to have these models adequately describe systems that undergo continuous macroscopic change as described by thermodynamics. Hence, what renders DP problematic is the discrepancy between two modes of description – BSM's models on the one hand and the thermodynamic description on the other. And this discrepancy becomes problematic in particular within the context of the foundational project. However, if no inter-theoretic reduction of thermodynamics on BSM is sought, and in particular, if no reduction of $T_{TD}$ on $T_{BSM}$ is sought, one might as well live happily with this discrepancy.

## The framework of BSM

Let's take a closer look at the framework of BSM as it is usually presented, e.g. in discussions of approach to equilibrium:[5] We are interested in a (classical) system comprised of $N$ constituents that move through physical space. The micro state of the system – a complete specification of the positions and momenta of all the constituents involved in the system – is represented by a point $x$ in its $6N$-dimensional phase space $\Gamma$. (If the energy of the system is constant, the phase space volume available to the system is reduced to a $(6N-1)$-dimensional energy hypersurface $\Gamma_E$.) The available phase space is partitioned into sub-regions that correspond to different macro states $M_i$ – specified by particular macro values taken on by macro variables, see above – of the system: the sub-region in $\Gamma$ corresponding to a macro state $M_i$ is the set of all $x \in \Gamma$ that instantiate $M_i$ and called the macro region $\Gamma_{M_i}$. Macro regions $\Gamma_{M_i}$ form a partition of $\Gamma$: they cover all of the available $\Gamma$ and do not overlap. A macro state of special interest is $M_{eq}$, the equilibrium state, instantiated by any micro state $x \in \Gamma_{M_{eq}}$ (and counterfactually instantiated[6] by all micro states $x \in \Gamma_{M_{eq}}$). Given the standard Lebesgue measure, the volume of sub-regions of $\Gamma$ can be assessed and hence is rendered comparable. It is usually said that (Frigg & Werndl 2012, 102) "for gases $\Gamma_{M_{eq}}$ is vastly larger [...] than any other macro-region, a fact also known as the 'dominance of the equilibrium macrostate' [...]; in fact $\Gamma_E$ is almost entirely taken up by equilibrium microstates. For this reason the equilibrium state has maximum entropy."

---

[5] Roughly following the presentation in Frigg & Werndl (2012, 101f.).

## Partitioning phase space into macro regions

Of special importance for the purposes of this paper is the notion of macro regions and their partitioning phase space (ibid.): "From a macroscopic perspective, the system is characterised by a set of *macrostates* $M_i$, $i = 1, ..., m$. To each macrostate corresponds a macro-region $\Gamma_{M_i}$ consisting of all $x \in \Gamma_E$ for which the system is in $M_i$. The $\Gamma_{M_i}$ form a partition of $\Gamma_E$, meaning that they do not overlap and jointly cover $\Gamma_E$." Frigg et al. (2016, 4.1) summarise the key feature of the correspondence between micro states and macro states in a short statement: "to every given microstate $x$ there corresponds *exactly one* macrostate".

For one thing, this means that $x$ either instantiates a certain macro state $M_i$ or another macro state $M_j$, $(i \neq j)$, but never both. Since macro states $M_i$ and $M_j$ correspond to macro regions $\Gamma_{M_i}$ and $\Gamma_{M_j}$, respectively, $x$ must not belong to two macro regions. This can be called the *no-overlap* condition:

$$\forall x \forall \Gamma_{M_{i,j}} \neg \left( x \in \Gamma_{M_i} \land x \in \Gamma_{M_j} \right); \; i \neq j \qquad \text{(no-overlap)}$$

For another, it means that macro regions exhaustively cover the available phase-space volume. All micro states must instantiate *exactly one* macro state, read as: every micro state must instantiate *some* macro state, and hence belong to *some* macro region. It cannot belong to *no* macro region at all. Hence, all of the available phase space must be covered by macro regions ("they … jointly cover $\Gamma_E$." (Frigg & Werndl 2012, 102.)) This can be called the *exhaustive-cover* condition:

$$\forall x \exists \Gamma_{M_i}(x \in \Gamma_{M_i}) \qquad (exhaustive\text{-}cover)$$

---

[6] "Counterfactually instantiated" in the sense that at any time, only one micro state is actualized, and thus only one micro state actually instantiates the macro state, but the actualization of the other micro states in the same macro region *would* instantiate the same macro state, if they *were* actualized instead.

Taken together, *no-overlap* and *exhaustive-cover* ensure that there is always some macro state instantiated, that this macro state is internally consistent, i.e., that a micro state never instantiates two (or more) distinct macro states at the same time, and that macro regions are not fuzzy: while it might be the case that there is some uncertainty or even complete ignorance about which macro state a given system is actually in – if only because the respective macro values haven't been measured/observed – one can always be sure that its actual micro state $x$ is located in *some* macro region or other, i.e. in one (*exhaustive-cover*) and only one (*no-overlap*) macro region $\Gamma_{M_i}$, and that thus, one and only one macro state is instantiated at any given time. After all, fuzzy boundaries would mean that the actual micro state does not always belong to a certain (i.e. to one and only one) macro region. Being located at the fuzzy boundary, $x$ could belong to $\Gamma_{M_i}$, to $\Gamma_{M_j}$, to neither, or to both. The latter option "both" is excluded by *no-overlap*. The option "neither" is excluded by *exhaustive-cover*. Hence, it must belong to either one of the former. In other words: boundaries of macro regions are sharp.

These considerations about the overlap and fuzziness of macro regions – or rather: these stipulations of their impossibility – are reflected in the schematic pictorial representations of phase-space partitioning we find, e.g., in Goldstein et al. (2019)[7]:

---

[7] I thank Sheldon Goldstein and the publisher of Annalen der Physik for the permission to reproduce this figure. It first appeared in Goldstein et al. (2017). Copyright Wiley-VCH Verlag GmbH & Co. KGaA. Reproduced with permission.

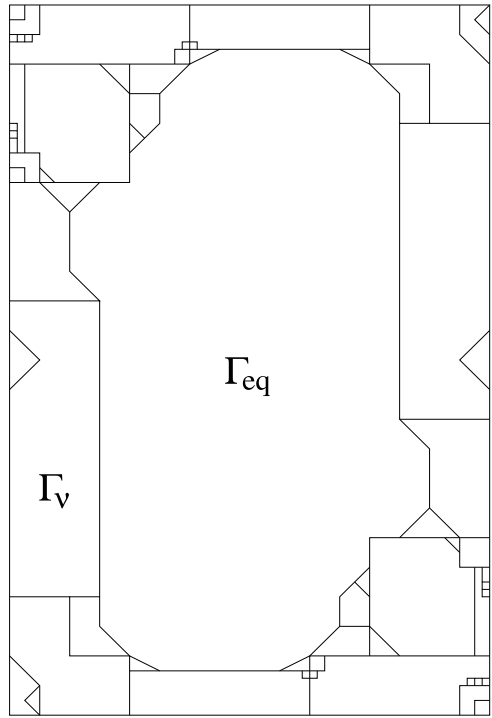

*Figure 1*

As can be seen, macro regions cover all of phase space, do not overlap and have sharp boundaries.

## The origin of the discontinuity problem

As stated above, macro regions in phase space are of finite, non-zero volume, and there is a finite number of macro regions in a partition. Why is that so? First, let's answer the question why macro regions must have a finite, non-zero volume. The reason for this lies in the way BSM works. The Boltzmann entropy function – $S_B(M_i) \coloneqq k_B \log[\mu(\Gamma_{M_i})]$[8] – relies on the volume of macro regions in order to assign different entropy values to different macro states of a system. If, contrary to this, macro regions in fact were of volume zero, the entropy function would assign to macro states no well-defined entropy value. This would seriously undermine the spirit and functionality of BSM. For example, it would be impossible to mark the equilibrium macro state as dominant in virtue of its having the highest entropy. Likewise, arguments about the relation of entropy change and the arrow of time would be blocked. This consequence is to be avoided, if only for the fact that it would be too high of a price to pay in comparison to what potentially could be gained by having zero volume macro regions.

---

[8] See Frigg & Werndl (2012, 102), Werndl & Frigg (2015, 1225).

In order to retain its spirit and functionality, BSM must construe macro regions as having finite volume greater zero.

Let's move on to the second part of the statement above: there is a finite number of macro regions in a partition. The reason for that is rather simple. Given macro regions, as construed within the framework of BSM, are of finite, non-zero volume (and do not approach zero volume), it is trivial that the finite volume of phase space available to a system can only be partitioned into a finite number of macro regions. But why even assume that the available phase space volume is finite? If it were infinite itself, it could be partitioned into an infinite number of compartments, despite the latter having a finite, non-zero volume. However, it is unreasonable to assume an infinite phase space volume for at least many of the systems BSM is supposed to model. For example, in the paradigmatic gas in a box, the spatial degrees of freedom each constituent has are bounded: particles can only move within the box (itself of finite volume). Such cases should be covered by BSM, so it is uncalled for to simply assume that BSM only models systems that are not limited to a finite spatial volume. Likewise, for finite temperatures, the momenta of the particles can only take on values within certain limits. Accordingly, the phase space volume of the entire system is restricted to be finite, if very large. And once the available phase space volume is finite, it is impossible to partition it into an infinite number of compartments with finite, non-zero volume. So the number of macro regions in a partition must be finite.

Having established that there is a finite number of macro regions in a partition, there can only be a finite number of different macro states. Macro states and macro regions, as described before, stand in a one-to-one correspondence. That is, the mapping between macro regions and macro states is bijective. But it is impossible to map a set containing an infinity of elements – let alone a set containing an uncountable infinity of elements – bijectively onto a set containing only a finite number of elements. Accordingly, it is impossible to bijectively map infinitely many macro states onto finitely many macro regions. So if there is a finite number of macro regions, they can only correspond to a finite number of different macro states. In fact, the number of macro states and the number of macro regions in the model must be equal: there are precisely as many macro states as there are macro regions, and this number is finite.

Given there is a finite number of different macro states, there can only be a finite number of different sets of macro values corresponding to them. In the BSM framework, a macro state, as defined above, is specified by a set of macro variables, taking on macro values. The simplest case, of course, is if this set only contains one relevant element, such that we can ignore the rest. Take, for example, some macro states that differ only in their temperature macro values. In this case, we can ignore other macro variables: they are assumed to take on identical values for the different macro states. Since there is a finite number of different macro states in the model, and these macro states differ in their temperature macro value, there is a finite number of different temperature macro values. The reason for this is similar to the one just discussed in the case of macro regions and macro states: the set of different macro states, since the latter are specified by temperature macro values, stands in a one-to-one correspondence to the set of different temperature macro values. So with a finite number of different macro states, there is a finite number of different temperature macro values specifying the former.

It is important to recognize that *different* macro regions must correspond to *different* macro states, and thus to *different* macro values. For if different macro regions would correspond to the same macro state, i.e. to identical macro values, they wouldn't be different macro regions in the first place. Such macro regions, corresponding to the same macro state, would be combined into one macro region. This is a consequence of the one-to-one correspondence between macro regions and macro states in the framework of BSM. In effect, different macro regions must correspond to different macro states, and, a fortiori, to different macro values. In the simplified example, being concerned only about the temperature macro variable, this means that different macro regions must correspond to different temperature macro values. But, as argued above, there is only a finite set of macro regions, and hence a finite set of different temperature macro values assigned to them. What, then, does it mean for a finite set of temperature macro values to consist of *different* temperature macro values? It means that, when subtracting any value from this set from any other, the result must always be different from zero, for if it was zero, the two values would be identical. It is unreasonable to assign identical macro states, i.e., macro states not differing in their macro values, to different macro regions, as argued before: those macro regions would be combined into one macro region, such that, in the end, macro regions and macro states remain in a one-to-one correspondence, and likewise macro regions and (sets) of macro values.

So there is a finite set of different macro values, each specifying a different macro state, corresponding to different macro regions. But if all pairs of elements of the set of macro values are different in that the subtraction of any one of them from any other results in a non-zero difference, then the transition between two macro values, i.e. between two macro states, i.e. between two macro regions, is always modelled as discontinuous. To illustrate this, imagine the temperature macro value of a system as modelled by BSM graphed as a function over time (figure 2):

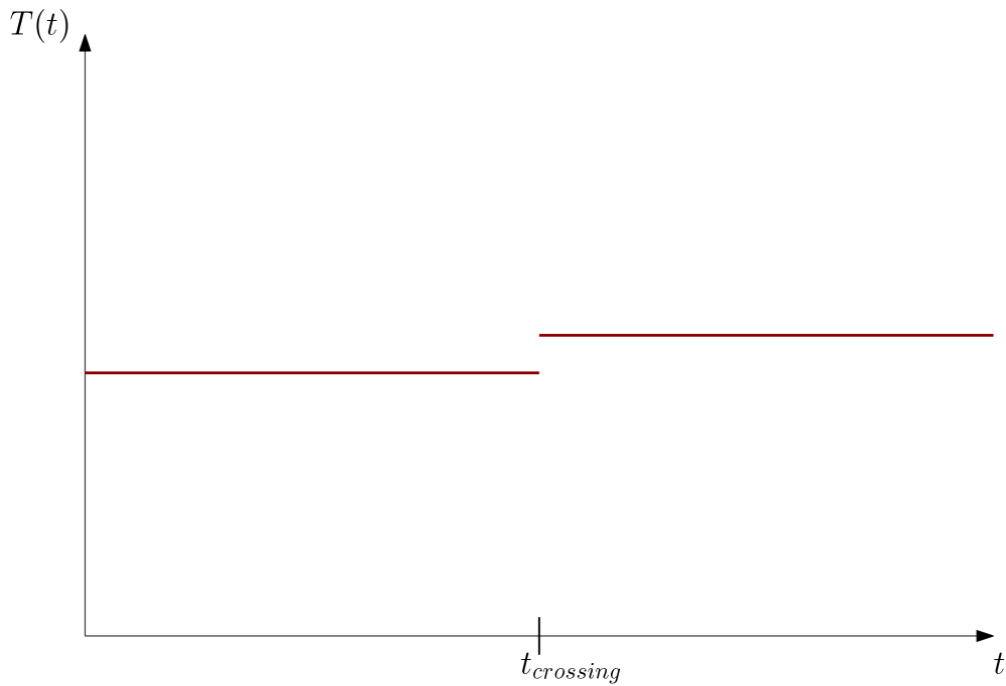

*Figure 2*

As long as the micro state wanders through a macro region $\Gamma_{M_i}$, the temperature macro value $T(t) = T_{M_i}$ remains absolutely constant, according to the model. So the graph shows a continuous, straight line parallel to the time axis, representing the constant temperature macro value. But at the precise instant the micro state crosses the boundary and moves into another macro region $\Gamma_{M_j}$ – call this instant $t_{crossing}$ – the temperature macro value changes to $T(t) = T_{M_j}$. Then, as long as the micro state stays in this macro region, the temperature macro value again remains absolutely constant, so the graph again shows a continuous, straight line parallel to the time axis, representing this different temperature macro value. Thus, the graphed function, at $t_{crossing}$, contains a discontinuity which is not removable: consider the limit $L^- = \lim_{t \to t_{crossing}^-} T(t)$ of the function for $t$ approaching $t_{crossing}$ from the left side, i.e. for the time during which the micro state moves through the first macro region $\Gamma_{M_i}$. The value of this limit is the temperature value before crossing the boundary: $L^- = T_{M_i}$. Now consider the limit $L^+ = \lim_{t \to t_{crossing}^+} T(t)$ of the function for $t$ approaching $t_{crossing}$ from the right side, i.e. for the time during which the micro state moves through the second macro region $\Gamma_{M_j}$, traced backwards. The value of this limit is the temperature value after crossing the boundary, $L^+ = T_{M_j}$. Since $T_{M_i} \neq T_{M_j}$, the limit from the left side and the limit from the right side are unequal as well: $L^- \neq L^+$. So there exists no single limit; the discontinuity at $t_{crossing}$ indeed is a jump discontinuity.

This is how the discontinuity problem arises from the usual framework of BSM. As stated before, DP consists in the discrepancy between the thermodynamic description of the time evolution of macro values, which can change continuously, and the fact that, in the usual framework of BSM, continuous change of macro values can't be modelled. So it neither consists solely in the fact that BSM is unable to model continuous change alone, nor does it contain the claim that in nature, or in thermodynamics, or in reality, macroscopic change is always discontinuous. The discontinuity encountered in BSM's models is an artefact resulting from the way the framework is set up: partitioning phase space into a finite number of macro regions, standing in a one-to-one correspondence with different macro states, the latter themselves standing in a one-to-one correspondence with different sets of macro values specifying them, such that, in the end, the elements of a finite set of macro regions stand in a one-to-one correspondence with the elements of a finite set of macro values – – this partitioning, judged from the thermodynamic viewpoint, where continuous change of macro values is allowable, is an idealisation. In which sense this idealisation is problematic will be discussed in the following.

## Models, fundamentality, and the discontinuity problem

The discontinuity problem of BSM is put into sharp relief in the context of the foundational project of reducing thermodynamics to BSM, regarding BSM as the fundamental theory, as for example Frigg & Werndl (2019) do. They classify BSM as a fundamental, true and complete theory (within the scope of statistical mechanics, and as opposed to Gibbsian statistical mechanics, which they consider to be an effective theory): "BSM is a fundamental theory" (425, 430, 436); "BSM provides a true description of the systems within the scope of SM" (425); "BSM provides the complete fundamental theory of SM systems" (431). By contrast to effective theories, fundamental theories are supposed to provide correct descriptions of the world (ibid. 431): "BSM is quite unlike GSM […]. Dynamical considerations occupy centre stage in BSM. It introduces macrostates with corresponding macroregions, and then defines equilibrium in explicitly dynamical terms (namely as the macrostate whose macroregion is such that, in the long run, the system's state spends most of its time in that macroregion). […] [U]nder the assumption that the world is governed by Newton's equation of motion […] the dynamics considered in BSM is the true dynamics at the fundamental level: the unabridged and unidealised dynamics with all interactions between all microconstituents of the system. Equilibrium results from macrostates that are defined in terms of macrovariables that supervene on the true microdynamics of the system, and where a system fluctuates away from equilibrium it does so as a result of the true underlying dynamics. In a classical world the theory gives a full account of all this – nothing is left out and nothing is averaged over. BSM provides the complete fundamental theory of SM systems."

Clearly, Frigg and Werndl engage in the foundational project, explaining macroscopic (thermodynamic) behaviour qua reduction to the more fundamental level as modelled by BSM. In order to make clear why this isn't as trivial as it seems, one first needs to distinguish between two types of functions models can have, and the way in which they can be deficient.

The first type of function of models is to "*represent* a selected part or aspect of the world, which is the model's target system." (Frigg & Hartmann 2020, 1., my emphasis.) Scale models have this function. They are (ibid.) "down-sized or enlarged copies of their target systems. A typical example is a small wooden car that is put into a wind tunnel to explore the actual car's aerodynamic properties." Whether a model fulfils its representational function successfully depends on the application. The wooden car model might be successful in representing the aerodynamic aspects of the real car, but it certainly fails in representing the real car's material composition (see ibid.). It is deficient in the sense that it doesn't represent all aspects of the real car. With respect to the intended application, this deficiency isn't problematic. There, its function is to represent a certain aspect of the real car, not all of them.

A different type of function of models is to *describe* a target system that falls within the domain of description of a given theory or theoretical framework. For example, a system of two massive objects orbiting each other in empty space is a target system that falls within the domain of description of Newtonian classical mechanics. Models describing a target system often entail approximations or idealisations. Borrowing the distinction from Norton (2012, 210), "an approximation is an inexact description of a target system." For example, using Newtonian mechanics to describe the target system of two objects orbiting each other in empty space results in a model ignoring relativistic effects. This amounts to an approximation. "An idealization", on the other hand, "is a real or fictitious system, distinct from the target system, some of whose properties provide an inexact description of some aspects of the target system" (ibid.). (Fictitiously) modelling the above two-body target system as a system of two point-like masses would be an idealization. This model isn't merely an approximation. It also has a representational function in that it represents the aspects "mass" and "position" of the target system in a fictitious system of point-like masses. Different functions of models aren't necessarily mutually exclusive. Approximations are deficient in virtue of being inexact, and the same is true for idealisations, for they entail approximations. Idealisations however, because they are systems themselves (if only fictitious ones), may also be deficient in virtue of not representing all aspects of the target system. Again, these deficiencies aren't problematic, as long as the model fulfils its intended function.

The function of BSM's models is of the second type. They describe some aspects of a target system, e.g., of a gas in a box, that falls within the domain of description of BSM. To be more precise, they describe the macro evolution of a thermodynamic target system, e.g., during approach to equilibrium. Due to its method of partitioning phase space, BSM's models are an idealisation in that they replace the thermodynamic target system with a fictitious system that can only instantiate a subset of the macro states the target system can assume.[9] Their representation of the time evolution of the target system, in other words, is deficient in virtue of not representing all aspects of it. This idealisation entails also an approximation: BSM's models ignore the actual macro values of the thermodynamic target system's macro variables at any given time and approximate them by assigning fixed macro values for finite intervals of time.

Now this *approximation* is unproblematic in practice. If the model is too crude, if the approximation is too inexact, one can always impose a more fine-grained partition. However, that BSM's models are *idealisations* with respect to the possible macro states a thermodynamic target system can take on *is* problematic for the foundational project. Working "under the assumption that the world is governed by Newton's equation of motion", Frigg & Werndl regard BSM as a "true complete fundamental theory". As Norton (2012) points out, one would expect the less fundamental theory to be an idealisation of the fundamental theory, not the other way around, as is the case here. Clearly, BSM's models being idealisations with respect to the possible macro states a system can take on is an oddity that must not be overlooked when engaging with the foundational project of reducing thermodynamics to BSM, especially if the latter is supposed to be not only a fundamental, but also a true and complete theory.

Partitioning phase space: a known issue

As described above, the construal of phase space partitioning within the framework of BSM, with sharp boundaries between a finite number of non-overlapping macro regions of finite, non-zero volume, underlies the discontinuity problem. At the same time, this construal seems to be standard. There are hints to be found in the literature, showing that not everyone considers it to be entirely unproblematic. Goldstein et al. (2019, 4) remark that phase space partitioning is not as straightforward a procedure as one might think: "$\Gamma(X)$ is the set of all phase points that 'look macroscopically the same' as $X$. Obviously, there is no unique precise definition for 'looking macroscopically the same,' so we have a certain freedom to make a reasonable choice". And later (ibid., 28), introducing the idea of "'fuzzy' macro sets", i.e. fuzzy macro regions: "The point here is to get rid of the sharp boundaries between [macro regions] as the boundaries are artificial and somewhat arbitrary anyway." I agree with Goldstein et al.'s verdict that boundaries are "artificial". They are an artefact, resulting from the way the framework of BSM is set up.[10]

---

[9] By contrast, Frigg & Werndl (2019, 431) are correct when they write that the dynamics of the micro state "is the true dynamics at the fundamental level: the unabridged and unidealised dynamics with all interactions between all microconstituents of the system", given "the assumption that the world is governed by Newton's equation of motion". Reduction has already taken place "in the background" (Sklar 1993, 348), in the sense that, as Sklar puts it (ibid.), "we have already identified gases as collections of interacting molecules and reduced our gas theory to a theory of molecules and their interaction." The threat to the foundational project doesn't consist in putting into question this background reduction. (More on that below.) That is, it doesn't stem from the ontology and micro dynamics assumed by BSM, which, in this context, i.e., given the above assumption, isn't an idealisation in Norton's sense. Rather, the threat exists due to discrepancies regarding macro properties and their evolution, and stems from BSM's method of partitioning phase space.

Here is Wallace (2018, 19) taking a similar line regarding the arbitrariness of phase space partitioning in BSM: "the macrostate partition at the heart of Boltzmannian statistical mechanics is […] vulnerable to these criticisms [(concerning the adequacy of coarse-graining)] […] Consider some standard descriptions of the coarse-graining: '[W]e must partition [phase space] into compartments such that all of the microstates $X$ in a compartment are macroscopically indistinguishable[.]' (Callender 1999, p.355). 'Everyday macroscopic human language (that is) carves the phase space of the universe up into chunks.' (Albert 2000, p.47) If pushed, I suspect Boltzmannians would reply that it is not the epistemic indistinguishability of macrostates that is doing the work, but rather the possibility of writing down robust higher-level dynamics in terms of macrostates, and largely abstracting over microscopic details". Wallace's argument here is a tu quoque: if the Gibbsian approach can rightly be criticized for coarse-graining, then the Boltzmannian approach can rightly be criticized for partitioning phase space into macro regions. The Gibbsian approach is not under consideration here, so this issue shall not further defer the discussion. The point to be made is merely that others see a potential for criticism of the BSM framework as well, and locate it at the same juncture: the partition of phase space into macro regions with sharp boundaries.[11]

---

[10] That they are artificial is not to say that they are necessarily "subjective" or "anthropocentric". There is a strong analogy between coarse-graining in GSM and partitioning in BSM. Robertson 2020 defends both against attacks referring to the alleged subjectivity and anthropocentricity, justified by what Robertson (op.cit. 564) calls (ibid.) "the MI justification" – the appeal "to our [limited] observational capacities". As she points out, "the MI justification is unsatisfactory". Most importantly, without going into too many details, coarse-graining, and by analogy, partitioning, are (op.cit, 565) "abstraction[s] to a higher level of description", a level of description that in its own right is endowed with explanatory power and describes regularities in the behaviour of macroscopic systems such as gases (see op.cit. 566). All this would be lost if it wasn't for coarse-graining or partitioning. As will become clear below, I strongly agree with Robertson's take: to simply adopt an infinitely "fine-grained" partition would amount to not partitioning at all. It would render BSM dysfunctional. That being said, for the issues discussed in this paper, it is irrelevant how the partition is justified, and whether this justification itself is justified or not. As long as we do partition phase space in the usual way, the issues discussed here arise, independently of the particular partition employed and its justification. I am indebted to an anonymous referee for the pointer to Robertson's account.

## The micro state crossing sharp boundaries

Let's make DP apprehensible and reveal its consequences in a series of examples. Imagine a system of interest, e.g., a gas in a (sealed, insulated) room. Modelling it in the usual way, a partition of its available phase space including sharp boundaries between macro regions is introduced. These macro regions correspond to different macro states of the gas, such that, if the micro state of the gas is located within a certain macro region, the gas instantiates the macro state as specified by its corresponding macro values. Hence, the actual micro state of the gas determines its actual macro state. Let's say that, in its initial macro state $M_0$ at $t_0$, the gas is confined to a certain corner of the room. Let the system evolve until the gas has spread out and evenly fills the whole room. This state is called $M_{eq}$ and marked with the time stamp $t_{eq}$. Clearly, the micro state representing the system must have evolved and crossed some boundaries. To simplify the picture, imagine that there are only two macro states: $M_0$ with the gas in the corner, and $M_{eq}$ with the gas spread out. Accordingly, there are two macro regions, $\Gamma_{M_0}$ and $\Gamma_{M_{eq}}$. Since the gas has changed its macro state from residing in the corner to filling the whole room, its micro state must have wandered from $\Gamma_{M_0}$ to $\Gamma_{M_{eq}}$. Hence, it must have crossed the boundary between these macro regions. Zooming in on the phase space region where the micro state crosses the boundary between $\Gamma_{M_0}$ and $\Gamma_{M_{eq}}$, there is a precise instant when this happens, namely when the micro state lies exactly on the boundary. Let's mark this instant $t_{crossing}$. Just before $t_{crossing}$, the micro state is in $\Gamma_{M_0}$, such that it is instantiating the macro state $M_0$, with the gas in the corner. Just after $t_{crossing}$, the micro state is in $\Gamma_{M_{eq}}$, such that it is instantiating the macro state $M_{eq}$, with the gas spread out. Hence, the continuous evolution of the micro state over just a tiny distance in phase space, such that it crosses the boundary, amounts to a rather dramatic discontinuity in the macro state of the gas in this model: at $t_{crossing}$, the volume jumps from a rather small value (the gas confined in the corner of the room) to a comparatively large value (the gas spread out over the entire room). Yet, judging from the viewpoint of thermodynamics, as well as from considerations about the speeds at which gas particles usually move – at finite speeds, certainly – one would expect a small, continuous evolution of the micro state only to amount to an accordingly small, and likewise continuous evolution of the macro state. After all, the continuous change of the micro state around $t_{crossing}$ might represent nothing but the displacement of one measly particle by a few microns. Such a tiny difference between micro states before and after $t_{crossing}$ can hardly account for the tremendous discontinuity between macro states $M_0$ and $M_{eq}$.

---

[11] The following exchange between Sheldon Goldstein and Daniel Sudarsky during the former's talk *Gibbs vs. Boltzmann Entropy* at the *Chimera of Entropy* summer school in Split in 2018 is also elucidating (my transcript):
Daniel Sudarsky: "Something that has bothered me all the time: consider that partition that you have [the usual partition with sharp boundaries]. Then consider a collection of points that are approaching the boundary from the two sides. I cannot imagine being able always to differentiate between them [... inaudible]".
Sheldon Goldstein: "You are right. So maybe you need to develop a better scheme which is more realistic, so you don't have sharp boundaries. [...] You are quite right; in a better world than ours, somebody will have done that. Maybe somebody has some kind of fuzzy notion of macro state. But you don't want to introduce that here [in the context of a heuristic discussion of entropy and approach to equilibrium], which will just complicate things. But you are completely right, there isn't that realistic division between [macro regions]: 'Oh, these points over here look like equilibrium, go epsilon further, doesn't look like equilibrium anymore.' That, obviously, is unrealistic. But if you want to get an understanding of the phenomena you should make that idealisation." (https://www.youtube.com/watch?v=jU838SXBrv4, 00:52:30 – 00:53:40)
Of course, such a statement, made in the context of a presentation, must not be mistaken for the "official" position of the speaker. That is, it must not be assumed that Goldstein (or Sudarsky) would have made the same or a similar statement in a published paper. (Neither, that they wouldn't make such statements in a published paper. After all, Goldstein et al. 2019 call a partition with sharp boundaries "artificial".) Nevertheless, this exchange can be regarded as a further hint to the effect that the way phase space is partitioned in the framework of BSM isn't unanimously viewed as entirely unproblematic.

One might want to reply that this strange correlation between the microscopic and macroscopic evolution of the system is due to the overly simplified example used. Certainly, there are not only two macro states, and, accordingly, not only two macro regions with one boundary between them. The partition is way too coarse. But irrespective of how many macro regions are introduced in accordance with equally many macro states, the boundaries between macro regions are always sharp, and macro regions are of finite number, corresponding to a finite number of different macro states of the gas. So, no matter how fine-grained the partition is, as long as it contains finitely many macro regions of non-zero volume, corresponding to finitely many different macro states, the problem will prevail: a small, continuous change in the micro state of the gas, while crossing the boundary, will amount to a discontinuous change in its macro state. The issue can be put like this: as soon as boundaries are introduced when carving up phase space, i.e., as soon as a partition into macro regions is introduced, this entails the assumption that micro states "left" and "right" of these boundaries, i.e. micro states in different macro regions, instantiate different macro states. Otherwise, it would be nonsensical to introduce this partition. But once there are sharp boundaries, DP arises, no matter how fine- or coarse-grained the partition. Differentiating between fewer or more macro states results in coarser or finer graining of the partition, but partitions will always be grainy, as long as macro regions are of finite, non-zero volume.

### Sharp boundaries being shifted

The following example exploits the aforementioned arbitrariness of the partition. Imagine a partition of the available phase space of a system. Two neighbouring macro regions, $\Gamma_{M_i}$ and $\Gamma_{M_j}$, correspond to different macro states $M_i$ and $M_j$ and are separated by a sharp boundary. Three micro states, $x_1, x_2, x_3$, are selected, such that $x_1$ lies somewhere in the middle of $\Gamma_{M_i}$, $x_2$ lies arbitrarily close to the boundary, but still in $\Gamma_{M_i}$, and $x_3$ resides equally close to the boundary, but opposite from $x_2$, in $\Gamma_{M_j}$. Figure 3 provides an illustration:

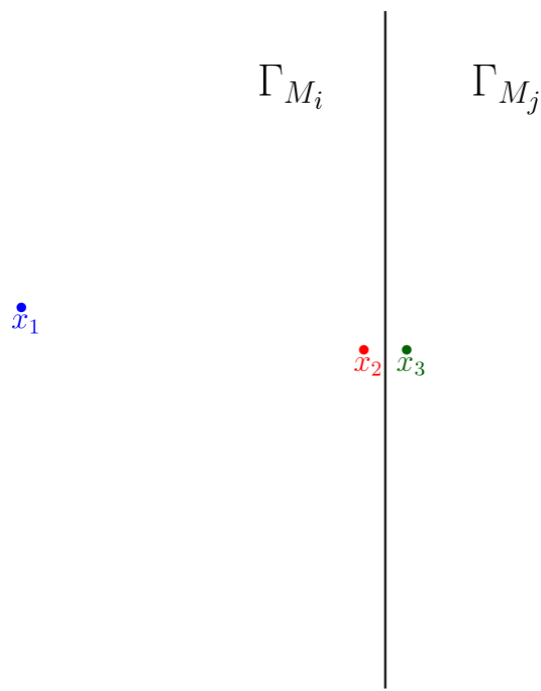

*Figure 3*

A tiny change of the partition, i.e. a small shift of the boundary between $\Gamma_{M_i}$ and $\Gamma_{M_j}$, should not amount to a model representing a completely different physical situation. Yet, this is what will happen. If the boundary indeed is drawn arbitrarily, it can be shifted, say, a wee bit towards $\Gamma_{M_i}$, such that the volume of $\Gamma_{M_i}$ decreases ever so slightly, while the volume of $\Gamma_{M_j}$ increases by the same amount. With $x_2$ lying arbitrarily close to the boundary, this shift results in both, $x_3$ and $x_2$, lying in $\Gamma_{M_j}$ (figure 4).

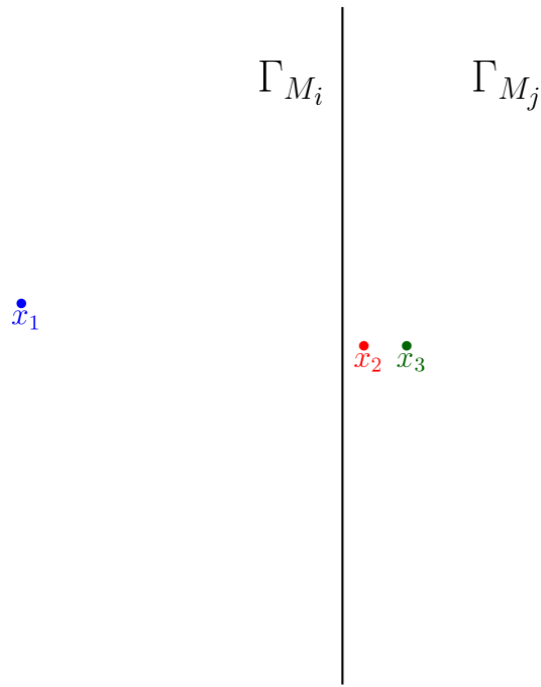

*Figure 4*

This new, slightly different partition seems an appropriate correction to the previous one, at least at first sight: the micro states $x_2$ and $x_3$, always residing extremely close together, are now grouped into the same macro region $\Gamma_{M_j}$, thereby, according to the model, instantiating the same macro state $M_j$. The relatively remote micro state $x_1$ remains in its original region $\Gamma_{M_i}$. From a macroscopic point of view, $x_2$ and $x_3$ are now "macro-identified" in that they are taken to instantiate the same macro state. This macro state is different from the macro state instantiated by the fairly remote $x_1$.

But all is not well. A tiny change of the partition, this example shows, amounts to models representing completely different physical situations on the macro level while nothing on the micro level changes: before the boundary shift, $x_1$ and $x_2$ were regarded as instantiating the same macro state $M_i$, while $x_3$ was instantiating $M_j$. After the shift, $x_2$ and $x_3$ are instantiating the same macro state $M_j$, while $x_1$ is still instantiating $M_i$. So in case $x_2$ becomes actual, it makes a huge difference in the model whether the boundary has been shifted or not. At the same time, nothing in the target system's micro state has actually changed: $x_2$ specifies the exact same micro state as before the shift of the boundary. The disagreement between the two models, despite describing the same microscopic situation of the same target system, is evident. Merely changing the partition slightly while keeping the actualised micro state fixed results in different macro states being instantiated. This suggests that, on BSM, the macro state doesn't depend on the micro state *alone*, but also on the partition, which can be chosen rather freely. This is a troubling result.

And there is more: $x_1$, $x_2$ and $x_3$ have been arbitrarily selected. They are just phase points, like any other. So why not select another micro state, call it $x_4$, also arbitrarily close to the (now shifted) boundary, opposite of $x_2$ (figure 5)?

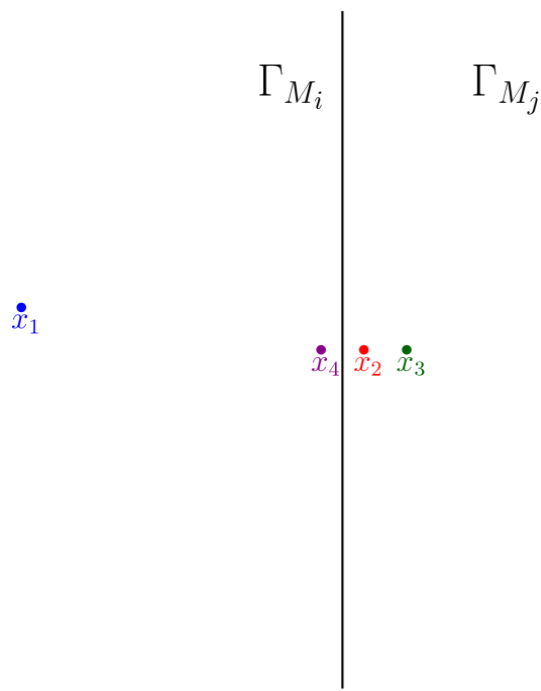

*Figure 5*

The same problem as before reappears. So the "correction" of the partition turns out to be no correction after all. Shifting boundaries around, in effect, is just shifting around a discontinuity; it is shifting the problem to a slightly different location, thereby making the underlying problem more evident.

Sharp boundaries between macro regions in phase space, as these examples should have illustrated, are just as artificial as the border between the Netherlands and Belgium.[12] There is a line, drawn arbitrarily, but in its immediate vicinity, on both sides, things look quite the same.

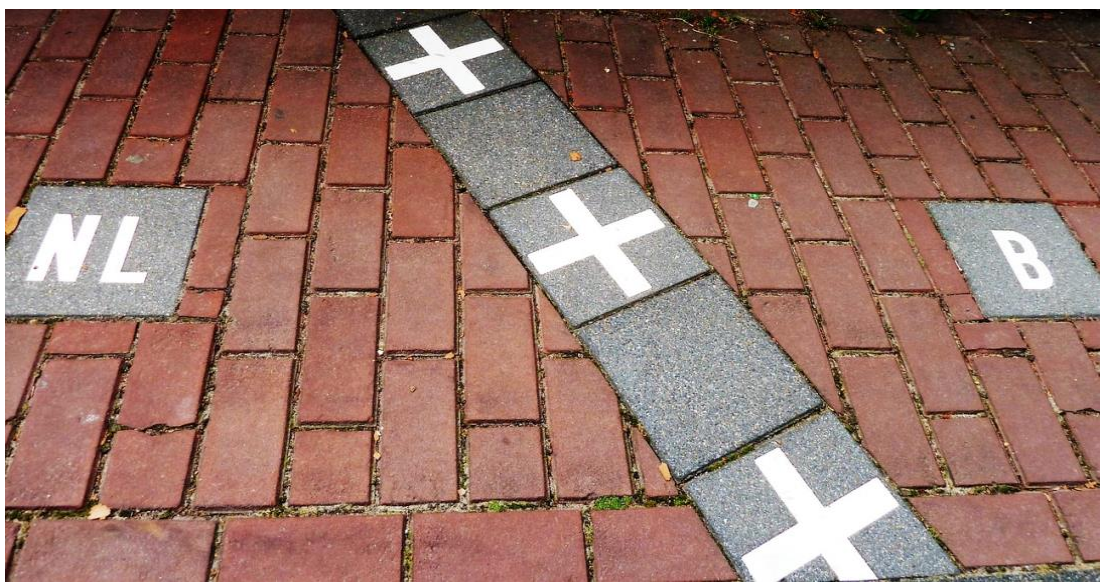

## The case of entropy

There is yet another, related issue arising from sharp boundaries that at least deserves mentioning. In the usual framework, the Boltzmann entropy function is generally defined as a function assigning an entropy value to macro states, depending on the volume of their respective macro regions: $S_B(M_i) \coloneqq k_B \log [\mu(\Gamma_{M_i})]$. The larger the volume, the higher the entropy. Thus, the largest region, i.e. the equilibrium region, is assigned the highest entropy. Accordingly, increase of Boltzmann entropy of a system[13] is described by the micro state moving to regions of larger volume (Goldstein et al. 2019, 19): "increase of Boltzmann entropy means that the phase point [$x$] moves to bigger and bigger macro sets [$\Gamma_{M_i}$]."

---

[12] Incidentally, the border between Belgium and the Netherlands has shifted as recently as 2018. See Stone (2018).

[13] Boltzmann entropy of a system is defined as follows by Werndl & Frigg (2015, 1225): "The Boltzmann entropy of a system at time $t$, $S_B(t)$, is the entropy of the system's macrostate at $t$: $S_B(t) \coloneqq S_B(M_{x(t)})$, where $x(t)$ is the system's microstate at $t$ and $M_{x(t)}$ is the macrostate supervening on $x(t)$." See also the almost identical definition in Frigg & Werndl (2012, 102).

In a partition, different macro regions usually have different volumes, such that the micro state, upon traversing the boundaries between them, instantiates a macro state of different entropy. As a result, entropy changes in a jumpy fashion, like other macro values, as discussed above: upon the micro state moving from a smaller macro region to a larger one, entropy jumps from a lower to a higher value. The entropy function becomes discontinuous. By contrast, it stays at a constant value while the micro state wanders through a macro region, not crossing any boundaries. This behaviour, like the discontinuous behaviour of other macro values in BSM's models, might come as a surprise. Wouldn't one expect there to be a continuum of allowable entropy values, and that, while moving from one value to another, the continuum between them is instantiated, just like one would expect other macro values to evolve continuously? At least, it doesn't seem implausible to say that entropy and its change *ideally* would be described by a continuous function instead of a function that assigns the same entropy value during one finite time interval – the time interval during which the micro state traverses a macro region – and then another, different value during the finite time interval of traversal of the micro state through another macro region, with discontinuous jumps at the boundaries. But this is how entropy behaves in the models as they result from the standard framework. Arguably, this way of modelling entropy change is at odds with the continuous macro state evolution assumed for the thermodynamic target systems classical BSM seeks to describe (and reduce). As before, the origin of the issue can be traced back to the method of partitioning phase space into a finite set of disjoint macro regions with sharp boundaries and non-zero volume.

## Modifying the framework?

It has been argued that phase space partitioning with sharp boundaries poses a problem for modelling continuous macroscopic change in the usual framework of BSM. The examples above shall now be revisited while not assuming *exhaustive-cover* or *no-overlap*. These conditions, together with the requirement that macro regions are of non-zero volume, result in sharp boundaries. Let's see if, upon excising them, one at a time, the discontinuity problem vanishes.

### Removing exhaustive-cover

First, *exhaustive-cover* shall be removed, thus allowing for models in which there are micro states that do not belong to a macro region: $\exists x \forall \Gamma_{M_i}(x \notin \Gamma_{M_i})$. Regarding the partition, this means that there are regions of phase space that are neither a macro region, nor belong to one. Figure 6 provides an illustration:

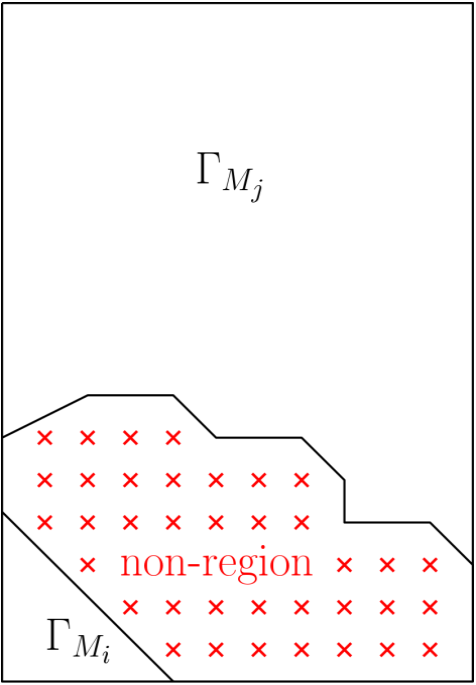

*Figure 6*

Via the correspondence between macro regions and macro states, phase points in those regions do not instantiate a macro state. Admittedly, a rather strange result. Under these conditions, it could well be that the micro state, while wandering through phase space, moves from a macro region to such a "non-region" and on to a macro region again, which would mean that, in between instantiating two macro states, for some time, it instantiates no macro state. But clearly, the constituents of the gas haven't lost their properties, all of a sudden. The phase points within such a non-region still describe the same microscopic features of the same system as the phase points in normal regions do: all the positions and momenta of all the system's constituents. Clearly, it is not the micro state of the gas that is responsible for this strangeness.

One might be tempted to resolve this issue by regarding these non-regions as transition regions between others. Recall the example of a gas in a room. The coarseness of this initial partition with only two macro regions can be reduced by introducing a "non-region" between these two that is taken to be a transition region. Let's call the resulting macro regions $\Gamma_{M_i}$, $\Gamma_{M_j}$ and $\Gamma_{M_{between}}$ (figure 7). That is all fine and well. The initial partition, after all, was too coarse. However, the so-called "non-region", then, is not a non-region, corresponding to no macro state at all. Instead, the points it contains instantiate a macro state $M_{between}$ between the initial state $M_i$ and the final state $M_j$. Taking this route, so-called non-regions are just ordinary regions, such that there are no true non-regions in phase space. *Exhaustive-cover*, in effect, isn't removed.

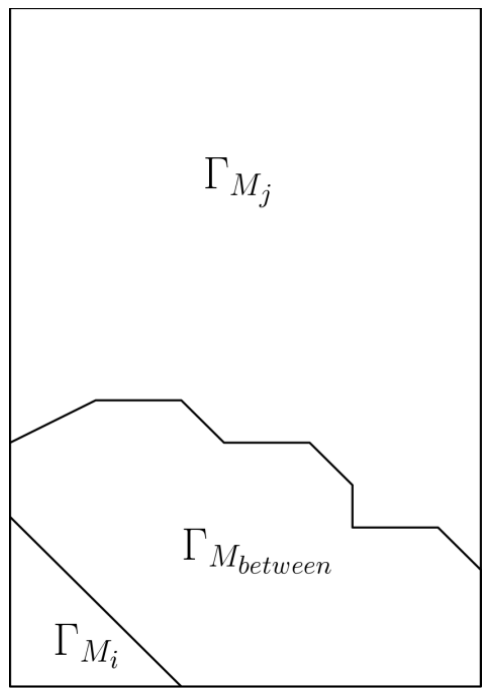

Figure 7

Furthermore, apart from the strangeness of having true non-regions in phase space, even if one regards them as a perfectly normal feature of BSM's models, one would be facing the same discontinuity problem as before, and even more drastically. In transitions from ordinary regions to ordinary regions, a jump discontinuity between macro values appears. In transitions from ordinary regions to non-regions and vice versa, an essential discontinuity appears: one of the limits $L^-$ or $L^+$ (see above) doesn't exist, since non-regions don't correspond to macro values. So there is not just a discontinuous jump between different macro values, but a jump from a macro value to an undefined value or vice versa, only exacerbating the issue.

As an interim result, it can be noted that it seems impossible to do away with the *exhaustive-cover* condition. Either, the resulting "non-regions" are true non-regions, making the discontinuity problem even worse, or they are ordinary regions after all. In any case, the original problem to be solved remains.

Removing no-overlap

So let's try keeping *exhaustive-cover* and excising *no-overlap* instead. This allows for micro states that belong to more than one macro region: $\exists x \exists \Gamma_{M_i} \exists \Gamma_{M_j} \left( x \in \Gamma_{M_i} \wedge x \in \Gamma_{M_j} \right); i \neq j$. Accordingly, there can be overlap regions in phase space (figure 8).

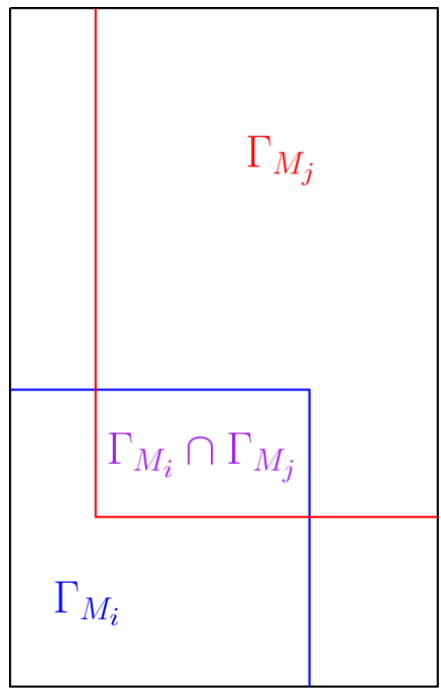

*Figure 8*

A micro state within such an overlap region $\Gamma_{M_i} \cap \Gamma_{M_j}$ is in $\Gamma_{M_i}$ as well as $\Gamma_{M_j}$, and therefore instantiates both, $M_i$ and $M_j$. As long as these have the same macro values, all is fine. However, they can't. As per usual, since macro states are specified via macro values, and correspond to macro regions, macro states being specified by the same macro values are identical macro states, i.e. there is only one macro state, corresponding to one macro region. In other words: macro regions corresponding to identical macro states form a union into one macro region. So, whenever one wants to speak of different macro regions, they must correspond to different macro states. And that, in turn, means that an overlap region of two different macro regions corresponds to two different macro states. Accordingly, every micro state in an overlap region instantiates two different macro states at the same time, which is an inconsistency.

To provide an extreme example of this equally strange result, recall again the simple gas example. Imagine that there is some overlap $\Gamma_{overlap} = \Gamma_{M_0} \cap \Gamma_{M_{eq}}$. If the micro state is in $\Gamma_{overlap}$, it is in $\Gamma_{M_0}$ and $\Gamma_{M_{eq}}$. Accordingly, the gas is concentrated in the corner ($M_0$) as well as evenly spread out over the whole room ($M_{eq}$) – an inconsistency even worse than DP. Hence, one should abstain from excising *no-overlap* as well.

## A different solution?

Removing either *exhaustive-cover* or *no-overlap* from BSM's framework results in severe problems. It thus seems reasonable to keep them. Yet, at the same time, leaving everything at status quo provides no solution to the discontinuity problem as it arises from sharp boundaries between a finite number of macro regions of non-zero volume. Consequently, one may ask whether a framework in which a set of phase points, i.e., a macro region, corresponds to *one and the same* macro state is a good one. Tweaking the framework's conditions didn't result in the desired solution. Maybe, then, it is time to consider adopting a different one.[14]

---

[14] Such an alternative, I present in a companion paper.

## The discontinuity problem and the foundational project

Let's finally dwell on the consequences DP has for the foundational project of reducing thermodynamics to BSM. As shown above, $T_{TD}$ and $T_{BSM}$ behave differently. This discrepancy lies at the heart of DP. But why is this a problem for the foundational project? Presumably the simplest and, as Sklar (1993, 340f.) writes, "one of the primary ways in which our theories of the world are unified" is "reduction by identification". "A general methodological principle seems to be that we ought to identify, as opposed to positing a correlation, whenever we can. That is, we should identify whenever the assertion of such an identification is not blocked by some feature of the situation, typically by the reduced entity having some feature genuinely not attributable to the reducing". When trying to reduce thermodynamics to BSM, one encounters such a blocking feature. The reduced theory (thermodynamics) has a feature, namely continuous change of macro values, that is not attributable to the reducing theory (BSM), which can handle only discontinuous change of macro values.

As Sklar points out, the reduction of thermodynamics to statistical mechanics *does* involve a genuine identification, insofar as it takes part in the broader reductive program (ibid., 341): "This is the reduction of the theory of macroscopic matter to its micro-constituents by the identification of the macroscopic entities as structured out of microscopic entities." BSM successfully does that. It takes macroscopic systems, e.g. gases, to be structured out of microscopic entities (ibid., 348): "the thermodynamics of gases can only be reduced to statistical mechanics after we have already identified gases as collections of interacting molecules and reduced our gas theory to a theory of molecules and their interaction." But he is also eager to point out that this is not the whole story – that the matter is more intricate (ibid., 341): "whether the reduction also involves, in any simple way, the identification of the thermal features of things (especially temperature and entropy) with features of things characterized at the reducing level in non-thermodynamic terms is a subtler matter." One could, for example, put into question whether the thermodynamic concept of temperature, which is reasonably attributed to macroscopic systems, can really be identified with the "kinematic-dynamic" (Sklar, 362) concept of average kinetic energy of the microscopic constituents in terms of their velocities and masses. At the very least, such a conceptual identification is certainly not trivial. But one doesn't even have to go this far in order to encounter a problem. The thermodynamic macroscopic temperature value of a target system ($T_{TD}$) can change continuously. The macroscopic temperature value of the BSM model ($T_{BSM}$) describing the target system cannot. Already these closely related concepts, both denoting temperature on the same, macroscopic level of description, are not identified trivially. It is easy to see this by asking: does $T_{TD}$ supervene on $T_{BSM}$? The answer is "no". In order to have $T_{TD}$ supervene on $T_{BSM}$, the latter must change whenever the former changes, which is not the case, due to DP. Supervenience is certainly necessary for identification, and a fortiori for reduction qua identification. So, since $T_{TD}$ doesn't supervene on $T_{BSM}$, reduction of the former to the latter qua identification is blocked. Adopting the terminology from Wimsatt (2006), one could say that the intra-level reduction from thermodynamics to BSM fails for the macro-quantity temperature.

This is quite unlike the reduction (qua identification) of the concept of a gas (pre kinetic gas theory) to the concept of a gas (post kinetic gas theory) in terms of a collection of interacting, particular constituents. The thermodynamically described, macroscopic gas clearly supervenes on its constituents: macroscopic, thermodynamically described change of the gas is always accompanied by microscopic, kinematic-dynamic change, as reflected by the fact that the micro state evolves through phase space. That is, whenever $T_{TD}$ changes, the micro state changes as well. But not so the BSM temperature macro value $T_{BSM}$. In BSM's models, $T_{BSM}$ remains fixed for a finite interval of time (during the micro state's evolution within a macro region) while, during the same time interval, the thermodynamic temperature value $T_{TD}$ changes. So $T_{BSM}$ doesn't always change whenever $T_{TD}$ changes. So $T_{TD}$ does not supervene on $T_{BSM}$. Instead, the supervenience relation holds the other way round: whenever $T_{BSM}$ changes, $T_{TD}$ changes as well. However, this is the wrong way for the foundational project. For a reduction of thermodynamics to BSM to be possible, $T_{TD}$ must supervene on $T_{BSM}$. In effect, the foundational project is partially successful, and partially not. It is successful in that thermodynamically described systems can be reduced to systems of interacting constituents qua identification. That is, *inter*-level reduction succeeds. It is unsuccessful in that macro properties as described by thermodynamics cannot be reduced to macro properties as described by the models of BSM, because, as the exposition of DP establishes, the former do not supervene on the latter. Hence, *intra*-level reduction fails.

## Does $T_{TD}$ supervene on $T_{BSM}$?

Above, I have argued that $T_{TD}$ cannot be reduced to $T_{BSM}$ because the former does not supervene on the latter. In this section, I want to make this failure of supervenience, and a fortiori of reduction, explicit in a more technical way.[15]

For the sake of the following argument, we restrain ourselves to the classical, kinetic gas theory.[16] Here, thermodynamic temperature $T_{TD}$ is conceptualized as a measure of the average kinetic energy $\overline{E_{kin}}$ of the constituents of a system, e.g., of the particles of an ideal, monatomic gas. The kinetic energy of a single particle is given as

$$E_{kin} = \frac{1}{2}mv^2$$

with $m$ denoting the mass of the particle and $v$ its velocity. Since in an ideal, monatomic gas $m$ is the same for all particles, the average kinetic energy of all particles of the gas is given as

$$\overline{E_{kin}} = \frac{1}{2}m\overline{v^2}$$

where the bars denote averages, i.e., $\overline{v^2}$ is the average squared velocity of the particles and, accordingly, $\overline{E_{kin}}$ their average kinetic energy.

With the above conceptualization of thermodynamic temperature as a measure of average kinetic energy, $T_{TD}$ is directly proportional to $\overline{E_{kin}}$:

$$T_{TD} \propto \overline{E_{kin}}$$

Via the ideal gas law, with $k_B$ denoting the Boltzmann constant, this proportionality is expressed as

$$\overline{E_{kin}} = \frac{1}{2}m\overline{v^2} = \frac{3}{2}k_B T_{TD}$$

---

[15] I am indebted to an anonymous referee who pressed me on spelling out this argument in greater technical detail.

[16] If one should be concerned about the assumption of kinetic gas theory and insist on construing gases as continua for thermodynamics, the same argument can be construed without referral to individual gas particles by imagining the arbitrarily small increase of the temperature of such a continuous gas by an arbitrarily small amount of energy transferred into it from an external system, as heat or work exerted on it. The main purpose of construing the argument based on kinetic gas theory not only for the BSM framework but also for thermodynamics is to emphasize that the problem of reduction discussed here is nothing to do with the underlying physical ontologies assumed for the respective theories. The argument gains clarity by assuming the same physical ontologies for both, but it goes through even without this assumption, since the intertheoretic reduction we are after is an issue of how the theoretical properties of the theories – the values of the variables with which they represent the phenomenon of macroscopic temperature and its change – relate to each other: whether they instantiate the appropriate supervenience relation or not.

i.e.

$$T_{TD} = \frac{m\overline{v^2}}{3k_B}$$

$\frac{m}{3k_B}$ is a positive, real-valued constant. For simplicity, let's call it $\kappa$, such that

$$T_{TD} = \kappa\overline{v^2}$$

This shows the direct correspondence between thermodynamic temperature and average particle velocity in the case of ideal, monatomic gases.

Imagine now that the thermodynamic temperature increases by some amount $\Delta T_{TD}$. With the above relation $T_{TD} = \kappa\overline{v^2}$, where $\kappa$ is constant, some corresponding increase of the average particle velocity squared must account for it. And with $\overline{E_{kin}} = \frac{1}{2}m\overline{v^2}$, this corresponds to an increase in average kinetic energy. Thus, whenever $T_{TD}$ changes, average kinetic energy changes as well. The former supervenes on the latter (figure 9).

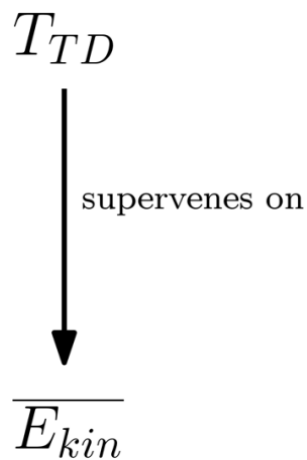

Figure 9

We can even show that this supervenience relation holds in the opposite direction, too. Imagine that the velocity of one particle is increased by some amount $\Delta v$ while everything else remains unchanged. Already intuitively, it is clear that this increase in velocity of one particle amounts to a proportional increase of $T_{TD}$:

$$\Delta v \propto \Delta T_{TD}$$

The average velocity (i.e., the arithmetic mean)[17] of all particles of the system at time $t_1$ is the sum of the velocities of the individual particles over the number $N$ of particles:

$$\overline{v_{t_1}} = \sum_{i=1}^{N} \frac{v_i}{N}$$

If at $t_2$ the velocity of one particle is increased by $\Delta v$, this increases to

$$\overline{v_{t_2}} = \left(\sum_{i=1}^{N} \frac{v_i}{N}\right) + \frac{\Delta v}{N}$$

Thus, the average velocity $\overline{v_{t_2}}$ at time $t_2$ is larger than the average velocity $\overline{v_{t_1}}$ at $t_1$ by $\frac{\Delta v}{N}$:

$$\overline{v_{t_2}} = \overline{v_{t_1}} + \frac{\Delta v}{N}$$

The average squared velocity of all particles of the system at time $t_1$ is the sum of the squared velocities of the individual particles at $t_1$ over the number $N$ of particles:

$$\overline{v_{t_1}^2} = \sum_{i=1}^{N} \frac{v_i^2}{N}$$

If the velocity of one particle – particle 1, say – is increased by $\Delta v_1$, this increases to

$$\overline{v_{t_2}^2} = \left(\sum_{i=1}^{N} \frac{v_i^2}{N}\right) + \frac{2v_1\Delta v_1 + \Delta v_1^2}{N} = \overline{v_{t_1}^2} + \frac{2v_1\Delta v_1 + \Delta v_1^2}{N}$$

When plugging this into

$$T_{TD} = \kappa \overline{v^2}$$

we see that $T_{TD2}$ must be larger than $T_{TD1}$, by the amount $\kappa\left(\frac{2v_1\Delta v_1 + \Delta v_1^2}{N}\right)$:

$$T_{TD2} = \kappa\overline{v_{t2}^2} = \kappa\left(\overline{v_{t1}^2} + \frac{2v_1\Delta v_1 + \Delta v_1^2}{N}\right) = \kappa\overline{v_{t1}^2} + \kappa\left(\frac{2v_1\Delta v_1 + \Delta v_1^2}{N}\right)$$

$$= T_{TD1} + \kappa\left(\frac{2v_1\Delta v_1 + \Delta v_1^2}{N}\right)$$

As long as $v_1$ and $\Delta v_1$ share the same sign ($v_1\Delta v_1 > 0$) and $\Delta v_1$ is finite, i.e., as long as $\Delta v_1$ is an actual increase of velocity in the same direction as $v_1$ (or in any direction, if $v_1$ is 0), it is always the case that

$$\kappa\left(\frac{2v_1\Delta v_1 + \Delta v_1^2}{N}\right) > 0$$

---

[17] Instead of the arithmetic mean one could use the root mean square (rms). However, the difference is irrelevant for the present argument; taking the rms would merely complicate calculations without any gain in clarity. See, e.g., Giancoli (2010, 630f.).

and thus that

$$T_{TD2} > T_{TD1}$$

This shows that, in thermodynamics, an increase in the velocity of one constituent particle of a system, with the other parameters held fixed, amounts to an increase in temperature of that system. Importantly, this $\Delta v$ can be arbitrarily small; as long as it isn't zero, thermodynamic temperature increases. Thus, whenever average kinetic energy changes, $T_{TD}$ changes as well. The former supervenes on the latter (figure 10).

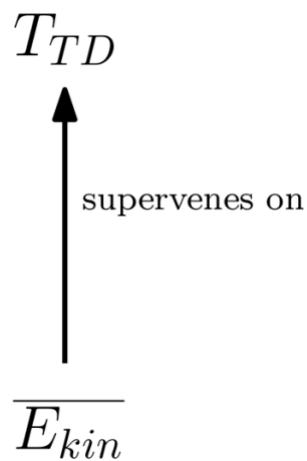

*Figure 10*

That $T_{TD}$ and average kinetic energy supervene on each other doesn't come as a surprise. This symmetric supervenience is a necessary condition for the background reduction qua identification already mentioned above. When temperature is conceptualized as a measure of average kinetic energy, it is only to be expected that the two supervene on each other.

How does this situation present itself in the BSM framework? There, the micro state of a system consisting of $N$ constituents is given as a point $x$ in a $6N$-dimensional phase space $\Gamma$, where the three position degrees of freedom and the three momentum degrees of freedom of each of the $N$ particles are represented by one dimension each. In other words, the micro state is represented by an ordered $6N$-tuple $(q_1, q_2, \ldots, q_{3N}, p_1, p_2, \ldots, p_{3N})$, where the $q_i$ each represent the position of one particle in one spatial dimension, and the $p_i$ each represent the momentum of one particle in one spatial direction. Since momentum is defined as the product of mass and velocity, $p = mv$, it can be increased by increasing velocity: $p + \Delta p = m(v + \Delta v)$. With $p = mv$, the increase of momentum is

$$\Delta p = m\Delta v$$

As before, suppose the velocity of one particle – particle 1 – is increased by $\Delta v$ in a certain direction, e.g., the second spatial direction, such that its momentum is increased by $\Delta p$ in that direction as well. That is, the micro state changes from

$x_1\,(q_1, q_2, \ldots, q_{3N}, p_1, p_2, \ldots, p_{3N})$ to $x_2\,(q_1, q_2, \ldots, q_{3N}, p_1, p_2 + \Delta p_2, \ldots, p_{3N})$.

Choose $\Delta v$ such that the resulting $\Delta p_2$ is so large that $x_2$ resides in a different macro region than $x_1$, i.e., that $x_2$ instantiates a different macro state than $x_1$ does. Then, $T_{BSM2}(x_2) \neq T_{BSM1}(x_1)$. In fact, whenever $T_{BSM2}(x_2) \neq T_{BSM1}(x_1)$, $x_2 \neq x_1$, because $T_{BSM1}$ and $T_{BSM2}$ correspond to different macro regions, and macro regions don't overlap, such that $x_1$ and $x_2$ cannot be identical (see above). Since the only difference between $x_1$ and $x_2$ is a difference in the velocity of one particle, $x_1$ and $x_2$ differ in average kinetic energy. Thus, whenever $T_{BSM}$ changes, average kinetic energy changes as well. The former supervenes on the latter (figure 11).

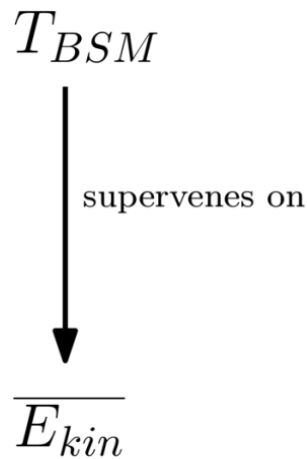

*Figure 11*

But now recall from above that $\Delta v$ can be chosen arbitrarily small. If $\Delta v$ can be arbitrarily small, then so can be $\Delta p$. Make it so small that the micro state $x$, after increasing the velocity of particle 1 by $\Delta v$ in the second spatial direction, remains in the same macro region in the BSM picture. I.e., $x_1$ and $x_2$ are both elements of the same macro region: $x_1, x_2 \in \Gamma_{M_i}$. But since all micro states in $\Gamma_{M_i}$ are assigned the same temperature macro value $T_{BSMi}$ via the one-to-one correspondence between macro regions and sets of macro values, $x_1$ and $x_2$, according to the BSM picture, instantiate the same temperature: $T_{BSM1}(x_1) = T_{BSM2}(x_2)$. Since the difference between $x_1$ and $x_2$ is a difference in the velocity of one particle, $x_1$ and $x_2$ differ in average kinetic energy. At the same time, $T_{BSM1}$ and $T_{BSM2}$ don't differ. Thus, average kinetic energy doesn't supervene on $T_{BSM}$.

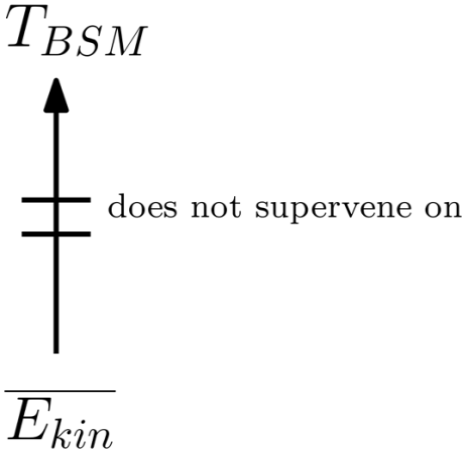

*Figure 12*

We have seen that, if $\Delta v$ is sufficiently small, $T_{BSM1}(x_1) = T_{BSM2}(x_2)$. At the same time, according to the thermodynamic picture, for the same $\Delta v$, $T_{TD2} > T_{TD1}$. Even though in both cases, the velocity of the same particle was increased in the same direction by the same amount $\Delta v$, thermodynamic temperature $T_{TD}$ did change while BSM-temperature $T_{BSM}$ did not. Thus, there are cases in which $T_{TD}$ changes while $T_{BSM}$ does not change. And since supervenience of *A* on *B* requires that, whenever *A* changes, *B* changes as well, $T_{TD}$ does not supervene on $T_{BSM}$ (figure 13).

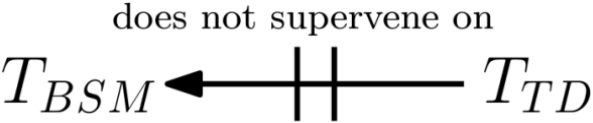

*Figure 13*

Conversely, $T_{BSM}$ does supervene on $T_{TD}$ (figure 14): whenever $\Delta v$ is chosen sufficiently large that $T_{BSM1} \neq T_{BSM2}$, it is also sufficiently large that $T_{TD1} \neq T_{TD2}$, because every finite $\Delta v$ suffices for that (see above).

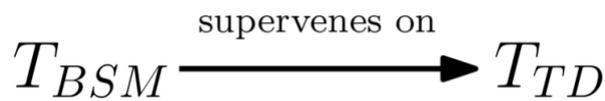
*Figure 14*

Let's put the pieces together. We have seen that a) $T_{BSM}$ supervenes on average kinetic energy, but not vice versa; b) $T_{TD}$ supervenes on average kinetic energy and vice versa; c) $T_{BSM}$ supervenes on $T_{TD}$, but not vice versa (figure 15).

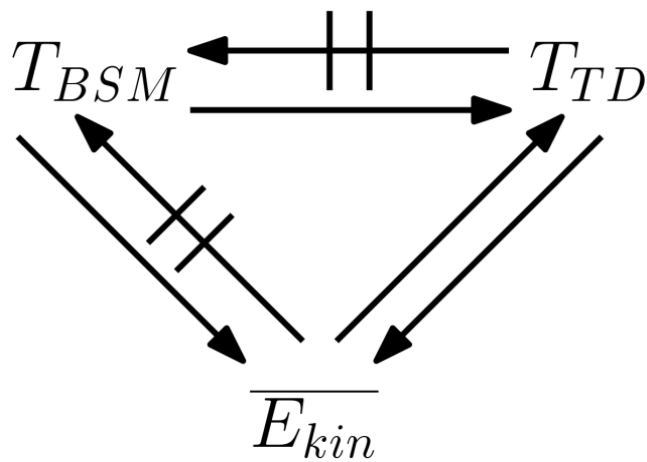
*Figure 15*

That $T_{TD}$ does not supervene on $T_{BSM}$ (but vice versa) is the most important result. It has been established by the independent arguments above, but it can also be seen when tracing the supervenience relations via average kinetic energy: $T_{BSM}$ does supervene on average kinetic energy, and the latter supervenes on $T_{TD}$. A fortiori, since supervenience is transitive, $T_{BSM}$ does supervene on $T_{TD}$. Conversely, while $T_{TD}$ does supervene on average kinetic energy, the latter does not supervene on $T_{BSM}$. Hence, $T_{TD}$ does not supervene on $T_{BSM}$.

One might be tempted to respond that the failure of supervenience of $T_{TD}$ on $T_{BSM}$ is all due to the specific way the partition was set up, and that partitioning into ever smaller macro regions would circumvent the issue. The first half of this response would be correct: indeed, the issue arises because of the partition. However, the second half of the statement is not correct: partitioning $\Gamma$ into ever smaller macro regions is of no avail, since $\Delta v$ can be chosen arbitrarily small. Without modifying the BSM framework substantially, the issue could be avoided only if one chooses to make the partition so small that, ultimately, every macro region contains only one micro state; that, however, amounts to the same as not partitioning at all. It is the same as directly assigning sets of macro values to individual phase points, which would be a nice thing to do, if we were able to do it. However, one of the main perks of SM is to circumvent the fact that we can't. On top of that, we'd be giving up advantages of the BSM framework, e.g. the possibility to define Boltzmann entropy as a measure of the size of macro regions: $S_B(M_i) \coloneqq k_B \log[\mu(\Gamma_{M_i})]$. With an infinitely fine-grained partition, not only would all macro states be assigned the same entropy, but that entropy value would be undefined – a situation that can hardly be satisfactory.

## Does thermodynamics reduce to BSM?

We have just seen that $T_{TD}$ does not supervene on $T_{BSM}$. As van Riel and Van Gulick (2019, 4.5.3) write, supervenience is a *conditio sine qua non* for reduction: "[Supervenience] surely is a necessary condition for reduction." McLaughlin & Bennett (2021, 3.3, my emphasis) even maintain that this conviction is held unanimously: "*Everyone* agrees that reduction requires supervenience." And they go on: "[O]n any reasonable view of reduction, if some set of *A*-properties reduces to a set of *B*-properties, there cannot be an *A*-difference without a *B*-difference."[18] In the case discussed here, which is a case of intertheoretic reduction, the sets

---

[18] An anonymous referee has pointed out that supervenience might not be a *conditio sine qua non* for *all* kinds of reduction. I concur; indeed, it is possible to challenge the universality expressed in these quotations. However, for the kind of reduction envisaged by the foundational project, I think that supervenience *is* necessary. To flesh this out with two examples, suppose that supervenience isn't necessary for reduction. Then, reduction is possible without supervenience. That is, it is then possible to reduce some A property to some B property without B changing whenever A changes. First, it is then possible to reduce the temperature of an ideal, monatomic gas to its micro state without the latter changing whenever the former changes. That, however, entails that the gas temperature can change without the micro state undergoing any change whatsoever. At least in my view, this is inconceivable: how would a physical macro property of a physical object change without the physical micro state of the *very same* object changing as well? Such a position would undermine the very basic assumption of physical composition: that macroscopic objects such as gases are made up of smaller stuff, atoms in this example, and that their properties and behaviour arise from the properties and behaviour of the smaller stuff. And because of that, it would undermine the background "reduction of the theory of macroscopic matter to its micro-constituents by the identification of the macroscopic entities as structured out of microscopic entities." (Sklar 1993, 341.) As Sklar points out, without this background reduction, the reduction of thermodynamics to statistical mechanics is impossible. Similarly, in the case of the temperature macro property: if supervenience wasn't necessary for reduction, one could reduce $T_{TD}$ to $T_{BSM}$ without the latter changing whenever the former changes. Surely, that would alleviate the problem discussed here, but it is questionable whether this approach is reasonable and in the spirit of the foundational project. How would a macro variable ($T_{TD}$), describing a physical macro property of a physical object, change, without another macro variable ($T_{BSM}$), intended to describe the *very same* physical macro property of the *very same* physical object, changing as well? As long as one engages in the foundational

of *A*- and *B*-properties mentioned by McLaughlin & Bennett are properties of the respective theoretical frameworks, i.e. thermodynamics and BSM. More precisely, they each are the theoretical elements that represent the temperature of a target system in their respective theoretical frameworks, the values of the temperature variables $T_{TD}$ and $T_{BSM}$, respectively. For thermodynamics to reduce to BSM, according to the statements just quoted, the value of $T_{BSM}$ should change whenever the value of $T_{TD}$ changes. But as we have seen, this is not always the case: $T_{TD}$ does not supervene on $T_{BSM}$. Thus, thermodynamics does not reduce to BSM.

With supervenience being a *conditio sine qua non* for reduction, any kind of reduction of thermodynamics on BSM is blocked. But one might want to elaborate a bit. I am willing to concede that BSM partially – but *only* partially! – reduces thermodynamics, if only in a way that is not exclusive to BSM: the particular way in which BSM reduces thermodynamics is available even without BSM. Thus, this reduction of thermodynamics isn't BSM's merit.

Van Riel & Van Gulick (2019, 2.1) distinguish between two main strands of reduction, *diachronic* and *synchronic* reduction. The former construes reduction, amongst being "a temporal affair", as a process of "replacement of one theory by another theory, such that one theory (the reducing one) becomes the successor of the reduced theory".

However, for one theory to become the successor of another in this sense, it is not sufficient that it comes after its predecessor, in the temporal sense. This is just one aspect of the successor relation. Usually, this relation also comprises, as Kemeny & Oppenheim (1956, 7) put it, "the replacement of an accepted theory (or body of theories) by a new theory (or body of theories) which is in some sense superior to it. Reduction is an improvement in this

---

project, i.e., as long as one aims to fully reduce TD to BSM, one must ensure that BSM's models adequately track those of TD. And in order to ensure that, in turn, $T_{BSM}$ must change whenever $T_{TD}$ changes. And that is nothing other than to say that $T_{TD}$ must supervene on $T_{BSM}$. In short: for the foundational project, I see no way around the supervenience condition.

That being said, it might be worthwhile to try to come up with a kind of reductionist account that i) is in the spirit of the foundational project and the reduction it envisages; ii) does not do away with very basic assumptions such as physical constitution; but iii) does not require supervenience. It can already be said that this kind of reduction, then, must adhere to some constraints. For example, it must not simply be a limiting reduction (see Palacios 2019 for an enlightening demarcation of limiting reduction from other kinds of reduction) where the size of macro regions goes to zero, for the reasons extensively discussed above. This, however, is a separate – and intricate – project that lies outside the scope of this paper.

sense." Van Riel & Van Gulick (2019, 2.1, my emphasis) elaborate on this: "[Diachronic reduction] is also often described as a sort of theory-improvement or scientific progress: particularly one in which *the laws of the prior theory apply to a proper subset of the cases covered by the laws of the succeeding theory*".

An example that gets cited for this kind of reduction is the alleged reduction of Newtonian dynamics of massive bodies to special relativity: the latter came not only after the former, it also improved on it, in the sense that it covered a wider range of phenomena, while still covering non-relativistic cases, i.e. cases in the low velocity limit, where $v \ll c$. Thus, Newtonian laws apply to a proper subset of cases of special relativity, and consequently, the second requirement for diachronic reduction, apart from temporal succession, is fulfilled, at least in this "narrow demonstration" (Fletcher 2019).

Whether this really suffices to make the case for a successful reduction is a matter of ongoing debate, see Fletcher (2019). But at least the supervenience requirement is fulfilled for the crucial variables. For example, whenever Newtonian momentum $p_{Newt} = mv$ changes, relativistic momentum $p_{Relat} = \frac{mv}{\sqrt{1-(v/c)^2}}$ changes as well. Thus, $p_{Newt}$ supervenes on $p_{Relat}$. At least, there is no issue of failed supervenience standing in the way of this (alleged) example of diachronic reduction.

The reduction of the concept of a gas (pre kinetic gas theory) to the concept of a gas (post kinetic gas theory) is an example of successful diachronic reduction, see above. What's more, it is also an example of successful intertheoretic reduction in the Nagelian sense.

The Nagelian model describes intertheoretic reduction as an explanation relation between two theories (see ibid., 2.2.1). Insofar explanation is demanded of successful, intertheoretic reduction, one can state that indeed, BSM explains, to some extent, thermodynamics, and, in this restricted sense, reduces it successfully. However, the qualification "to some extent" is important here, because it actually is the "background reduction" of "our gas theory to a theory of molecules and their interaction" (Sklar 1993, 348) that does all the work, qua some identity assumptions: in the exemplary case of a gas, the spatial distribution of its constituent particles is identical to the (macroscopic) volume $V$ of the gas and thus explains it; the force per area the particles exert on the walls of a container is identical to the (macroscopic) pressure $P$ of the gas in that container and thus explains it; the average kinetic energy of the particles is identical to the (macroscopic) temperature $T$ of the gas and thus explains it; the transfer of kinetic energy from particles of one system to those of another, and with it the redistribution of particle velocities, is identical to conductive heat flow from a hotter to a colder body after they are brought into contact, and thus explains it; etc. The background assumptions relevant for these explanations are part and parcel of kinetic gas theory. But kinetic gas theory, while it also features in BSM, can be had without BSM. And the other crucial part of BSM, its partitioning of the available phase space of a system, doesn't contribute to this explanation. In fact, it introduces difficulties: even while gases are identified with collections of particles in BSM, and macroscopic temperature $T$ is thought of as average kinetic energy of these particles, BSM-temperature $T_{BSM}$, qua BSM's association of macro values with macro regions, merely approximates $T$, unless the partition is so fine-grained that it becomes dysfunctional and pointless. So, while a background assumption of BSM does all the reductive work, and a fortiori, BSM in this restricted sense does reduce thermodynamics, the complete BSM framework is not an integral part of this reduction, and $T_{BSM}$ may not be identified with temperature in the sense of average kinetic energy, although it supervenes on it. Thus, BSM, as opposed to the kinetic gas theory, does not reduce temperature qua identification. Hence the caveat.

Most importantly, however, this background reduction is not the reduction we are after. Our question was whether $T_{TD}$ reduces to $T_{BSM}$. And while both, $T_{TD}$ and $T_{BSM}$, supervene on average kinetic energy, via their common background assumption of kinetic gas theory, $T_{TD}$ does not supervene on $T_{BSM}$, and hence cannot be reduced to it. In other words, while prima facie one might be led to believe that the notion of temperature is the same in thermodynamics and BSM, it is, in fact, not.

And there is more. On the Nagelian account of reduction (and successors of it, e.g. the Generalised Nagel-Schaffner Model of Reduction, see Dizadji-Bahmani et al., 2010), explanation of the reduced theory by the reducing theory stems from derivability: the laws of the (corrected) reduced theory must be derivable from the laws of the reducing theory plus "some auxiliary assumptions" and bridge laws in order to explain and thus reduce it.

The law of the reduced theory we are concerned with here is the (corrected) law governing the time evolution of $T_{TD}$; the law of the reducing theory is the law governing the time evolution of $T_{BSM}$. So, the question arises: can the (corrected) law governing the time evolution of $T_{TD}$ be derived from the law governing the time evolution of $T_{BSM}$ (plus bridge law and auxiliary assumptions)? At least for a "corrected" version of thermodynamics, it would be unnecessary to recover the exact, uncorrected thermodynamic treatment from BSM. Unfortunately, this derivation is not possible. On the one hand, the only candidate auxiliary assumption that would render it possible is an infinitely fine-grained partition. But, as I have argued, this route is not available, in particular because it leaves BSM dysfunctional and would amount to the same as not partitioning at all, such that we are left with the description of the micro state and its time evolution, disregarding macroscopic regularities (see fn. 10 above). But the law governing the time evolution of $T_{TD}$ is not derivable from the law governing the time evolution of the micro state alone, without a proper partition, not even only in approximation. If it were, we wouldn't need BSM in the first place. On the other hand, when applying the usual partition, at least we have a candidate bridge law, something that looks like a bridge law, but unfortunately, isn't one. Dizadji-Bahmani et al. even provide the reason why there is no bridge law establishing the relevant correlation between $T_{TD}$ and $T_{BSM}$. They write (399. I have replaced their placeholders for theoretical terms, $t_P$ and $t_F$, by the ones that are relevant here, $T_{TD}$ and $T_{BSM}$): "A bridge law is a statement to the effect that (1) $T_{TD}$ applies if, and only if, $T_{BSM}$ applies, and (2) $\tau_{T_{TD}} = f(\tau_{T_{BSM}})$", where $\tau_{T_{TD}}$ is the value of $T_{TD}$ and $\tau_{T_{BSM}}$ the value of $T_{BSM}$. $f(\tau_{T_{BSM}})$ is a function that transforms $\tau_{T_{BSM}}$ into $\tau_{T_{TD}}$, i.e., the BSM temperature value into the thermodynamic temperature value. However, in the case we are discussing here, $f$ is not even a function, because uncountably many $\tau_{T_{TD}}$ correspond to every $\tau_{T_{BSM}}$. Hence $f(\tau_{T_{BSM}})$ does not map each element of $\tau_{T_{BSM}}$ (its domain) to exactly one element of $\tau_{T_{TD}}$ (its codomain). The second requirement for the bridge law isn't met, and thus, there is no bridge law connecting $T_{TD}$ and $T_{BSM}$ in the necessary way, even if the two are coextensive. Note that this has nothing to do with multiple realisability, which states that a property of the reduced theory may correspond to several properties of the reducing theory. Here, it is the other way round: several properties of the reduced theory correspond to one property of the reducing theory, which means that the reduced theory does not supervene on the reducing theory. Again, the lack of the appropriate supervenience relation between $T_{TD}$ and $T_{BSM}$ is the culprit.

For the present purposes, this exposition of the threat DP poses for the foundational project must suffice. In a companion paper, I will suggest an alternative method of carving up phase space that avoids DP while keeping the functionality and spirit of BSM intact. By avoiding DP, supervenience of $T_{TD}$ on $T_{BSM}$ will no longer be blocked and a fortiori, this threat to the foundational project can be avoided.

## Acknowledgements

I gratefully acknowledge financial support of this research by 1) the Research Grants Council of the University Grants Committee of Hong Kong through the Hong Kong PhD Fellowship Scheme and 2) the Federal Ministry of Education, Science and Research of Austria through the Ernst Mach Grant (worldwide). The funding sources have no involvement in the writing of this paper and the decision to submit it for publication.

## Declarations

The author declares that there are no conflicts of interest.

## References

Albert, D. Z. (2000). Time and Chance. Cambridge, MA: Harvard University Press.

Batterman, R. (2020). Intertheory Relations in Physics. In: Zalta, E.N. (Ed.), *The Stanford Encyclopedia of Philosophy* (Fall 2020 Edition).
https://plato.stanford.edu/archives/fall2020/entries/physics-interrelate/

Ben-Menahem, Y., & Hemmo, M. (Eds.) (2012). Probability in Physics. Springer Science & Business Media.

Callender, C. (1999). Reducing Thermodynamics to Statistical Mechanics: The Case of Entropy. The Journal of Philosophy, 96(7), 348–373.

Dizadji-Bahmani, F., Frigg, R., & Hartmann, S. (2010). Who's Afraid of Nagelian Reduction? Erkenntnis, 73(3), 393–412. http://doi.org/10.1007/s10670-010-9239-x

Fletcher, S. C. (2019). On the reduction of general relativity to Newtonian gravitation. *Studies in History and Philosophy of Modern Physics*, *68*, 1–15.
http://doi.org/10.1016/j.shpsb.2019.04.005

Frigg, R. & Hartmann, S. (2020). Models in Science. In: Zalta, E. N. (Ed.), *The Stanford Encyclopedia of Philosophy* (Spring 2020 Edition).
https://plato.stanford.edu/archives/spr2020/entries/models-science/

Frigg, R., & Werndl, C. (2012). A New Approach to the Approach to Equilibrium. In: Ben-Menahem, Y. & Hemmo, M. (Eds.) (2012). *Probability in Physics.* Springer Science & Business Media. 99-113.

Frigg, R., & Werndl, C. (2019). Statistical Mechanics: A Tale of Two Theories. *The Monist*, *102*(4), 424–438. http://doi.org/10.1093/monist/onz018

Frigg, R., Berkovitz, J., & Kronz, F. (2016). The Ergodic Hierarchy. In: Zalta, E. N. (Ed.), *The Stanford Encyclopedia of Philosophy* (Summer 2016 Edition).
https://plato.stanford.edu/archives/sum2016/entries/ergodic-hierarchy

Giancoli, Douglas C. (2010). Physik, 3$^{rd}$ edition. Pearson.

Goldstein, S., Huse, D. A., Lebowitz, J. L., & Tumulka, R. (2017). Macroscopic and microscopic thermal equilibrium. *Annalen der Physik*, 529, 1600301.
https://doi.org/10.1002/andp.201600301

Goldstein, S., Lebowitz, J. L., Tumulka, R., & Zanghì, N. (2019). Gibbs and Boltzmann Entropy in Classical and Quantum Mechanics. *arXiv.org*. https://arxiv.org/abs/1903.11870v2

Hemmo, M., & Shenker, O. R. (2012). The Road to Maxwell's Demon. Cambridge University Press.

Kemeny, J. G., & Oppenheim, P. (1956). On Reduction. *Philosophical Studies*, *7*(1), 6–19. http://doi.org/10.1007/BF02333288

McLaughlin, B. and Bennett, K. (2018). Supervenience. In: Zalta, E. N. (Ed.), *The Stanford Encyclopedia of Philosophy* (Winter 2018 Edition). https://plato.stanford.edu/archives/win2018/entries/supervenience/

Norton, J. D. (2012). Approximation and Idealization: Why the Difference Matters. *Philosophy of Science*, *79*(2), 207–232. http://doi.org/10.1086/664746

Palacios, P. (2019). Phase Transitions: A Challenge for Intertheoretic Reduction? *Philosophy of Science, 86*(4), 612-640. doi:10.1086/704974

Robertson, K. (2020). Asymmetry, Abstraction, and Autonomy: Justifying Coarse-Graining in Statistical Mechanics. *The British Journal for the Philosophy of Science*, *71*(2), 547–579. http://doi.org/10.1093/bjps/axy020

Sklar, L. (1993). Physics and Chance: Philosophical Issues in the Foundations of Statistical Mechanics. Cambridge University Press.

Stone, J. (2018). Belgium and the Netherlands swap land to change their national borders. https://www.independent.co.uk/news/world/europe/belgium-netherlands-national-border-change-meuse-vise-eijsden-maastricht-a8141166.html Last accessed December 6, 2020.

van Riel, R. and Van Gulick, R. (2019). Scientific Reduction. In: Zalta, E. N. (Ed.), *The Stanford Encyclopedia of Philosophy* (Spring 2019 Edition).
https://plato.stanford.edu/archives/spr2019/entries/scientific-reduction/

Wallace, D. (2018). The Necessity of Gibbsian Statistical Mechanics. In submission.
https://dornsife.usc.edu/assets/sites/1045/docs/gibbsboltzmann.pdf

Werndl, C. & Frigg, R. (2015). Rethinking Boltzmannian Equilibrium. *Philosophy of Science*, *82*(5), 1224–1235. http://doi.org/10.1086/683649

Wimsatt, W. C. (2006). Reductionism and its heuristics: Making methodological reductionism honest. *Synthese*, *151*(3), 445–475. http://doi.org/10.1007/s11229-006-9017-0